**Oxygen Reduction Reaction and X-ray Photoelectron Spectroscopy of Sputtered Fe-N-C Films**

Yun Xu[a], Michael J. Dzara[b], Sadia Kabir[a], Svitlana Pylypenko[a,b*], Kenneth Neyerlin[a*], Andriy Zakutayev[a,*]

[a] Materials and Chemical Science and Technology Directorate, National Renewable Energy Laboratory, Golden, CO 80401, USA.

[b] Chemistry Department, Colorado School of Mines, Golden, CO 80401

[*]*Corresponding authors:* Andriy.Zakutayev@nrel.gov, Kenneth.Neyerlin@nrel.gov, spylypen@mines.edu

Electrocatalysts for the oxygen reduction reaction (ORR) based on complexes of iron and nitrogen in a carbon matrix (Fe-N-C) are a promising alternative to platinum group metal (PGM) based catalysts in polymer electrolyte membrane (PEM) fuel cells. Further improvements of Fe-N-C catalysts would benefit from model thin film studies of activity and stability of catalytic sites, but synthesis of Fe-N-C model thin films is challenging. Here we report on synthesis and characterization of Fe-N-C thin films produced by co-sputtering iron and carbon in a reactive nitrogen atmosphere onto removable glassy carbon rotating disk electrode (RDE) tips. Scanning electron microscopy (SEM) measurements indicate that the Fe-N-C films deposited at high temperature are smoother than the films annealed at high temperature. ORR activity measured on the thin Fe-N-C films is greater for both high-temperature samples than for the room-temperature sample. From the analysis of X-ray photoelectron spectroscopy (XPS) data, exposure of the films to high temperatures results in increased graphitization of the carbon with the Fe-N-C films, and increased relative amount of graphitic and hydrogenated nitrogen species. Overall the results of this study demonstrate the feasibility of a thin film model system approach for studying active sites in PGM-free catalysts.

# 1. Introduction

Electrochemical energy storage and conversion devices are a cornerstone of a renewable, emission free energy sector. In devices such as metal-air batteries and polymer electrolyte membrane (PEM) fuel cells the oxygen reduction reaction (ORR) taking place at the cathode plays a crucial role in determining the overall performance and cost of the system; similar to the oxygen evolution reaction (OER) in PEM electrolyzers used for hydrogen generation. Since the ORR is a kinetically sluggish process, a significant amount of catalyst - typically based on platinum-group-metal (PGM) - is required, impeding the commercialization of fuel cell devices due to the disproportionately high cost of PGM catalysts.[1] Thus, extensive efforts have been devoted to reducing or eliminating the use of precious metals in the past decades, and the pursuit of low-PGM and PGM-free catalysts has become an extensively explored research area.[2-5]. A class of PGM-free electrocatalysts for the ORR consisting of nitrogen coordinated transition metals (TM-$N_x$, where TM = Fe, Co, Ni) embedded in a porous graphitic carbon matrix, (TM-N-C), have emerged as promising candidates.[3, 6-9] Recent advances have demonstrated PGM-free Fe-N-C catalysts with ORR activities on par with PGM-based catalysts in acidic media.[10][11] Most Fe-N-C catalysts are synthesized using a high temperature pyrolysis of metal, nitrogen, and carbon precursors followed by subsequent acid leaching of inactive metal/metal oxide species, resulting in multiple active functionalities and defect sites for the ORR.[12-14] Significant amounts of research have been dedicated to optimizing the performance and improving electrochemical kinetics of PGM-free electrocatalysts, but the role played by various active site moieties is still not completely understood.

Improvement of PGM-free catalysts could be significantly accelerated by understanding which Fe-N-C active sites promote a direct 4 $e^-$ ORR mechanism vs. a stepwise 2x2 $e^-$ ORR pathway, but there is still significant controversy surrounding this topic in the literature. Species with atomically dispersed iron coordinated with nitrogen defects within a carbon network are typically viewed as the most active sites, however debate continues as to the relative activity of other species, depending on factors such as degree of coordination (Fe ligated by 2, 3, or 4 nitrogen atoms), location of the Fe-$N_x$ complex in the carbon, and the electronic structure of the iron atom.[15-17] Identification of active Fe-$N_x$ is further complicated due to the presence of numerous types of N-C species that may also contribute to ORR activity. Some investigations link ORR activity with pyridinic nitrogen and hydrogenated nitrogen species[18], while others attribute enhanced ORR activity to the presence of graphitic or quaternary nitrogen moieties.[18, 19] Also, some mechanistic studies concluded that these nitrogen species are responsible for the partial reduction of oxygen to hydrogen peroxide through the indirect 2x2 $e^-$ mechanism, whereas others have proposed the direct 4 $e^-$ reduction of oxygen to water.[20, 21] Yet another group of reports have suggested that while nitrogen defects are necessary to create an active catalyst, it is the carbon atom adjacent to the nitrogen defect that is the true active site.[20, 22]

Understanding of the catalysts which have a heterogenous structure with numerous types of possible defects and locations of nitrogen sites is limited by information depth and spatial resolution of common characterization techniques. For example, X-ray photoelectron spectroscopy (XPS) of the N 1s core level has an information depth of 5-10 nm which can lead to significant contribution from species in the "bulk" of the catalyst powders that may be inaccessible for the ORR. In addition, the typical area of analysis in XPS is 0.5-5.0 mm, which means analysis provides information about species present over a large area and quantification provides heavily averaged data. Sputtered thin films are a promising model system to study ORR active sites, in conjunction with XPS, because of their potentially smooth and homogenous surface. One such notable prior study synthesized Fe-N-C thin films as model ORR electrocatalysts by ambient temperature magnetron sputtering followed by high temperature annealing in a mixed $Ar/N_2$ atmosphere.[23] Thin film combinatorial sample libraries of $Fe_xC_{1-x-y}N_y$ (0< x< 0.06, 0< y <0.5) were investigated, and the nitrogen and iron content of as-produced films were characterized by XPS. The films annealed at 800 °C showed the highest ORR activity, and had the highest initial N content and highest retained Fe content after acid exposure. However, due to the existence of Fe in as-deposited thin films, carbon nanotubes were observed to grow on the surface at these high temperatures (800 -1000 °C) and increased the surface area to a very large extent. Therefore, conclusions about the role of specific ORR sites may have been convoluted with surface morphological differences due to the presence of nanotubes.

The objective of this study is to explore controlled synthesis of Fe-N-C thin films with active surface area similar to the apparent surface area, and to correlate the kinetic ORR current with the surface chemistry. As shown in Figure 1, sputter deposited Fe-N-C materials were characterized using scanning electron microscopy (SEM) to determine the morphology of as-prepared samples, while RDE in acidic media was used to evaluate their ORR activity, and XPS was employed to understand chemical speciation of samples before and after RDE testing. The RDE results indicate that the Fe-N-C thin films processed at high temperature have higher ORR activity than at ambient temperature. SEM images show that heating the substrate during the deposition leads to smoother morphology than post-deposition annealing. XPS results demonstrate that sputtering methods can produce graphitic and N-containing surface species, but alongside remove Fe-containing species that have low stability under acidic ORR conditions. Overall, these results demonstrate the feasibility of thin film model system studies of ORR catalysts. Further improvements to maximize active site density and decrease the concentration of unstable species should be possible by tuning of the sputter synthesis parameters.

## 2. Experimental Methods

Both glassy carbon flat substrates and removable glassy carbon RDE tips were placed on the stationary sample holder, with iron and carbon targets on either side of the deposition chamber at 45-degree angle with respect to the substrates (Figure 1). A 2-inch diameter target of carbon (graphite, 0.25-inch thickness Kurt J. Lesker, 99.9% purity) and 2-inch diameter iron (0.125-inch thickness, Kurt J. Lesker, 99.999% purity) were sputtered using Radio Frequency (RF) power of 90 W and 15 W respectively. Nitrogen was introduced through an RF plasma source with a power of 250 W. Samples were sputtered in an Ar/$N_2$ (1:1) gas mixture with flow rate of 10 sccm for both Ar and $N_2$, in a chamber with a base pressure of $1 \times 10^{-6}$ Torr, and under process pressure of 15 mTorr. A total deposition time of 60 mins was used in order to produce an approximate film thickness of 50 nm. Three samples were prepared, including one deposited at room temperature (RT), one deposited with the substrate heated to 650 °C (HT), and one deposited at room temperature and post-annealed at 750 °C (AN). For the AN sample, the substrate was heated to 750 °C at a rate of 5 °C/min and held at 750 °C for 5 mins.

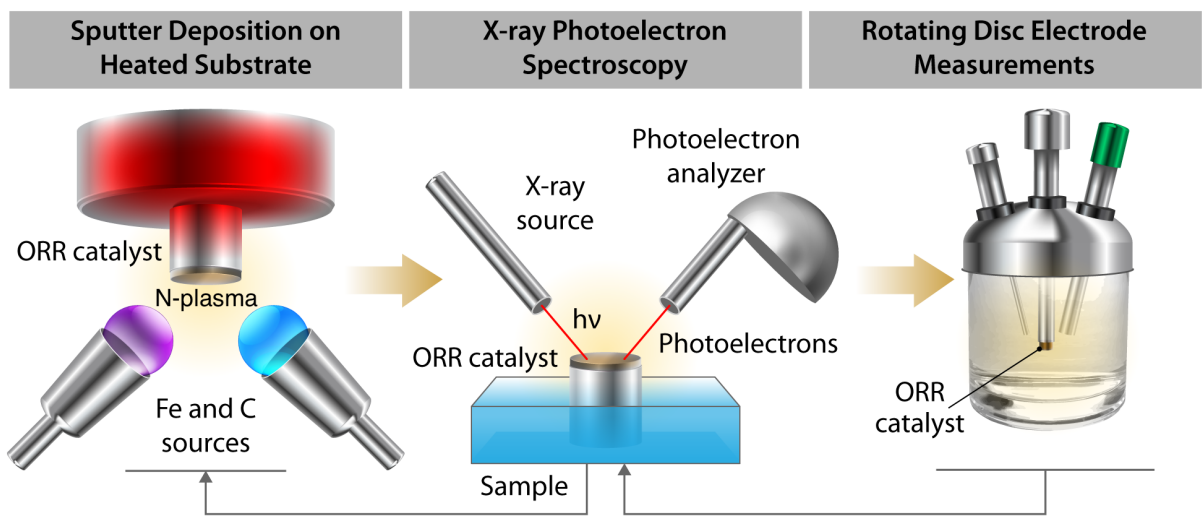

Figure 1: Illustration of methods flow of this study, including co-sputtering on heated substrate from Fe and C targets in N plasma, x-ray photoelectron spectroscopy before and after electrochemical measurements, and rotating disc electrode electrochemical measurements.

Electrochemical studies of the magnetron sputtered thin films were performed using the Pine Instrument Company electrochemical analysis system in a three-electrode cell containing 0.5 M $H_2SO_4$ electrolyte (Figure 1). A carbon counter-electrode and a Hg/$HgSO_4$ reference electrode was used. The working electrode was a glassy carbon rotating disc electrode (RDE) with a geometric area of 0.1967 cm$^2$.

The Fe-N-C sputtered thin films were deposited on the RDE to evaluate their electrocatalytic properties. The electrolyte was first saturated with $N_2$ for obtaining cyclic voltammograms (CV), first at a higher scan rate of 50 mV/s for stabilization, followed by slower scan rate at 5 mV/s. The gases were switched and the electrolyte was then saturated with $O_2$ for obtaining Linear Scan Voltammograms (LSV). Both the CV's and LSV's were obtained at 900 RPM and at 5 mV/s. The disc current densities were normalized to the geometric area of the glassy carbon disc electrode. All potentials (E) in the manuscript are referred to the reversible hydrogen electrode (RHE). The RHE potentials were confirmed by standardizing the $Hg/HgSO_4$ reference electrode at room temperature by immersing a platinum counter electrode into a solution of 0.5 M $H_2SO_4$ saturated with hydrogen and recording the voltage measurements between the counter and reference electrode. Following electrochemical testing, RDE tips were rinsed with deionized water and removed from the electrode housing for XPS analysis post-ORR testing. The morphology of each sample was determined using a JEOL JSM-7000F Field Emission Scanning Electron Microscope (FE-SEM). Images were taken at a 15.0 kV accelerating voltage and a working distance of 10 mm.

In order to quantitatively and qualitatively evaluate the as-produced and post-ORR thin film compositions, XPS was performed with a Kratos Axis spectrometer using a monochromatic Al $K_\alpha$ source operating at 300 W (Figure 1). The operating pressure was ~$2 \times 10^{-9}$ Torr, and the photon energy 1486.6 eV. Survey and high-resolution spectra were acquired at pass energies of 160 eV and 20 eV, respectively. Samples were mounted on conductive carbon tape so that charge neutralization was not necessary. Typically, three spots on each sample were measured, and any quantification reported is an average of those values. High-resolution spectra were recorded for the C 1s, O 1s, N 1s, and Fe 2p. XPS data was processed (CasaXPS software) using a linear background subtraction for quantification of C 1s, O 1s, and N 1s, and a Shirley background for the Fe 2p. Peak fitting of the N 1s was performed by a least-squares method using a series of components with a 70% Gaussian, 30% Lorentzian line shape. The same fit was applied to all samples, in which the full-width at half-maximum (FWHM) of each component was fully constrained and the position was allowed to vary by only 0.1 eV.

## 3. Results and Discussion

### 3.1 Oxygen Reduction Reaction and morphology

Prior to ORR measurements, sample morphology and surface area were studied using SEM. Figure 2 displays SEM images of the three samples featured in this study. From these SEM images, we can see that the RT sample has a smooth surface compared to the samples deposited or post-annealed at high temperatures. The AN sample treated at elevated temperature demonstrates the presence of rather large particles (estimated 100-200 nm in size), which also correlates with changes in the chemical state of iron species, as discussed later in the XPS section. Under the same magnification, the HT sample showed a more homogenous surface morphology compared to the AN sample, with estimated particle size of 30-50 nm.

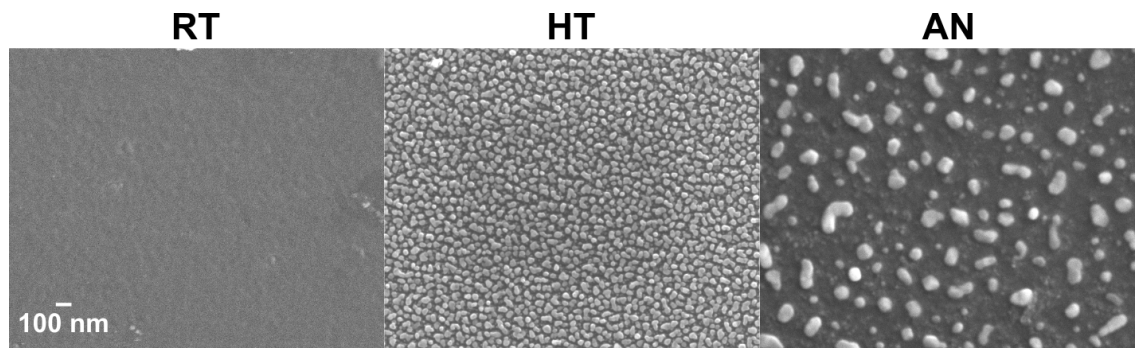

Figure 2. SEM images of each film are displayed, with images taken at the same x50k magnification. Samples deposited at high temperature (HT) show rougher morphology compared to samples deposited at room temperature (RT), but smoother morphology compared to RT-deposited samples that have been annealed (AN).

The ORR activity of Fe-N-C thin film model catalysts was evaluated using a RDE setup in acidic electrolyte. Figure 3a shows the CV results performed as a part of the stabilization procedure before LSV measurements for the Fe-N-C thin film model catalysts deposited on RDE tips. The CVs, taken in $N_2$-saturated acidic electrolyte, show the formation of a well-defined redox peak at 0.7 V vs. RHE which appears with increasing pyrolysis temperature. This peak is most defined for the AN sample, while no redox peak is observed for the RT sample. This redox peak has previously been attributed to the $Fe^{2+/3+}$-N redox transition in pyrolyzed Fe–N–C materials.[24]

Figure 3a further shows that the CV of the RT sample is also tilted and strongly deviates from a square signal, typical for samples with low electronic conductivity. Small peaks corresponding to the reduction and oxidation of some surface adsorbed species appear at 0.2 V and 0.3 V vs. RHE respectively. However, the pyrolyzed HT and AN samples have a higher capacitive signal, and defined redox peaks at 0.7 V vs. RHE. This suggests that thermal annealing not only has an effect on the formation of $Fe-N_x$ moieties, but also leads to the formation new nitrogen-containing surface species that are only formed at elevated temperatures.

Figure 3b shows the LSVs obtained in $O_2$ saturated electrolyte for the Fe-N-C thin film model catalysts. From the current densities obtained at 3 different potentials (0.2-0.4 V vs. RHE) from the LSV polarization curves in Figure 3b, it can be noted that the AN and HT samples outperform the RT sample, with the latter showing only negligible current densities. Moreover, the AN sample has a higher ORR onset potential, with a positive shift in potential by almost 150 mV compared to HT, demonstrating improvements in the ORR kinetics.

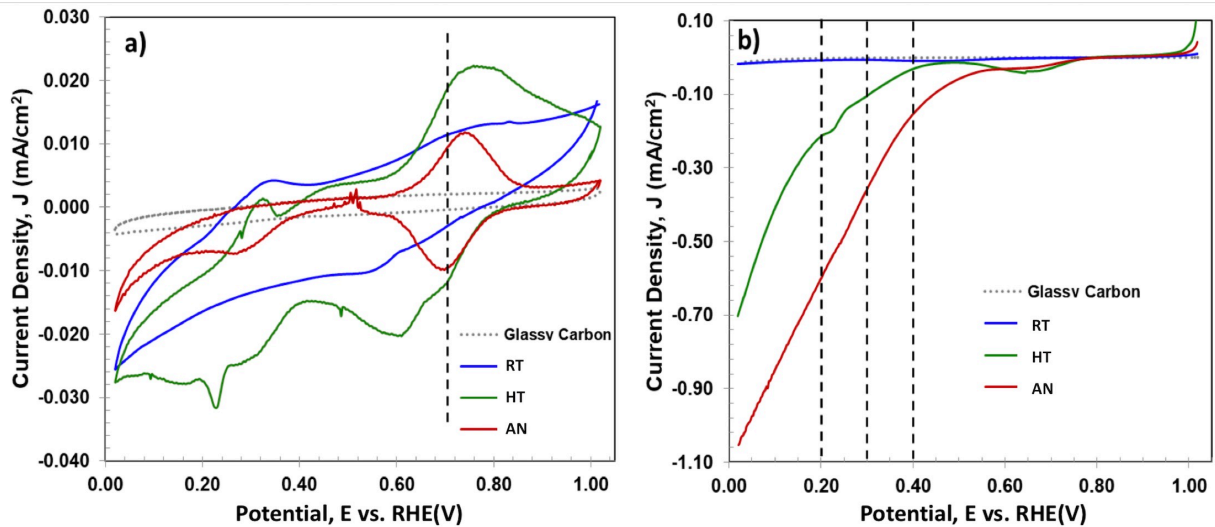

Figure 3. a) Cyclic Voltammograms (CV) and b) Linear Sweep Voltammograms (LSV) of the thin films. The HT and AN samples show strong CV redox peaks and are active ORR catalysts, in contrast to the RT sample.

Two additional important factors that affect the ORR performance of a catalyst are the number of active sites and pore structure of the catalyst layer. Corresponding performance of a material towards the ORR can be determined by the half wave potentials in the kinetic region. However, since these thin films have a thickness of 50 nm and are deposited on a non-porous glassy carbon, they were not able to attain diffusion limiting currents under potentials <0.65 V vs. RHE. This was also reported in other studies utilizing thin film model catalysts.[22]

Overall, the RDE results suggest that the Fe-N-C films processed at high temperatures (HT, AN) are active ORR catalysts, whereas the film deposited at room temperature (RT) is not ORR-active (Figure 3b). The deposition of thin films at room temperature might form some Fe-N-C moieties in the RT sample, but the majority of iron species most likely exist in forms that are not active for ORR or stable in acidic media, as suggested from the lack of $Fe^{2+/3+}$ redox peaks. Thus, counterintuitively the RT sample produced the lowest current densities, despite having the highest nitrogen abundance, determined using N1s XPS spectra as summarized in Table 1 and discussed next. This result motivates a more thorough investigation of the surface chemistry, and its evolution following RDE testing.

*3.2 X-ray Photoelectron Spectroscopy Analysis*

Quantitative compositional XPS analysis was performed on all films (both as-prepared and post-RDE testing), with the results summarized in Table 1. Temperature used during deposition or during annealing step after room-temperature synthesis has a major effect on nitrogen and iron speciation. First, post-RDE testing, iron content decreased below the detection level of XPS for all three samples, suggesting that all films contain primarily unstable iron-containing species. Indeed, many Fe-N-C synthetic schemes include acid leaching steps in order to remove inactive iron.[14] In the case of the films reported here, this step was performed *in situ* during electrochemical conditioning prior to collecting CVs and LSVs. Second, there is quite a large loss of nitrogen in the as-prepared HT and AN films relative to the RT film, accompanied by an increase in oxygen content. In contrast, post-RDE the trend in relative nitrogen content is reversed compared to the as-produced films – the RT film has the smallest amount of nitrogen, while the two films exposed to higher temperature show a higher nitrogen amount. This suggests that HT and AN films contain the most stable nitrogen species, despite having less nitrogen in the as-deposited films. However, we note that comparing relative compositional changes before and after ORR testing is convoluted with the decrease in iron content. While the decrease in iron content explains in part the changes in nitrogen film composition after ORR testing, the dramatic change seen when comparing the trend in nitrogen amount before and after ORR testing cannot be fully attributed solely to iron species dissolution.

Table 1. XPS compositional analysis of the films pre and post RDE testing. The Fe content is below the detection limit post-RDE testing. Pre-RDE testing the N content is higher for RT compared to HT and AN, but this trend is reversed post-RDE testing.

| Thin Film Samples | XPS: Pre-RDE Testing (at. %) | | | | XPS: Post-RDE Testing (at. %) | | | |
|---|---|---|---|---|---|---|---|---|
| | C 1s | O 1s | N 1s | Fe | C 1s | O 1s | N 1s | Fe |
| RT | 52.8 | 25.8 | 18.4 | 3.0 | 82.9 | 15.0 | 2.2 | <0.1 |
| HT | 61.7 | 29.3 | 3.4 | 5.6 | 68.3 | 25.6 | 6.1 | <0.1 |
| AN | 62.2 | 31.6 | 2.4 | 3.8 | 56.8 | 33.6 | 9.6 | <0.1 |

To get a deeper understanding of the surface chemistry of the studied Fe-N-C thin films, high-resolution XPS of each core level is analyzed pre and post RDE testing to identify possible trends in the active species present in the films. The N 1s is analyzed in detail through curve fitting (Figure 4), for which the development of the N 1s fit and assignment to chemical species is based upon the literature and on our previous experience fitting N 1s spectra.[16, 17, 25-27] A discussion of the rationale for the assignment of components to chemical species, and a table of fitting parameters is presented in Table S1, while a complete analysis of all other core levels (Fe 2p, O 1s, and C 1s) and a full discussion of the trends in surface chemistry with preparation conditions is found in Figure S1

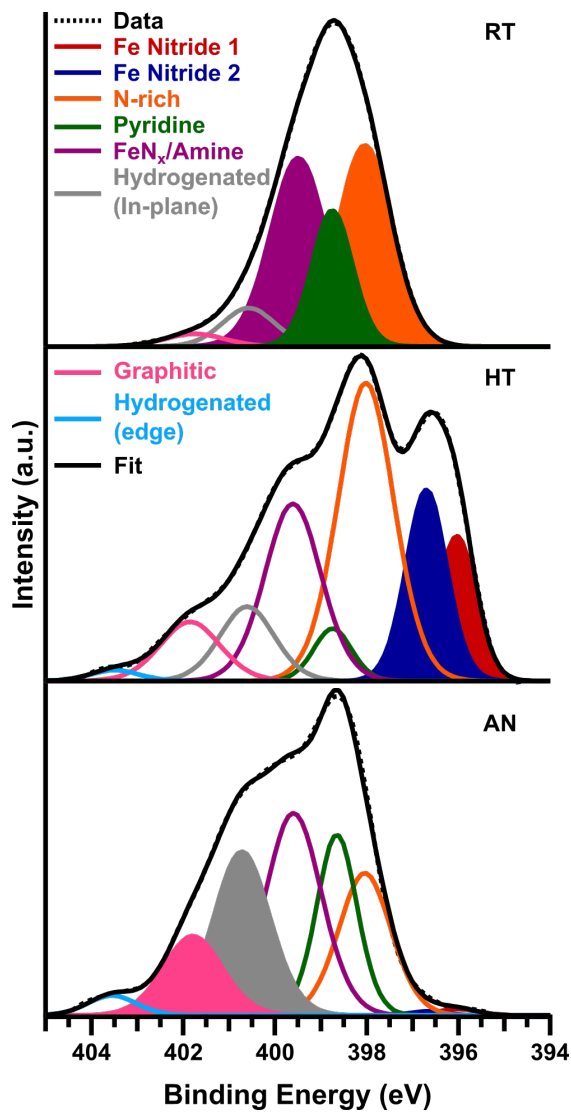

Figure 4. Representative curve fit, background-subtracted N 1s spectra are shown for all samples as-prepared. The RT samples show N in multiple species in C matrix, whereas for HT samples Fe-N species appear, and for AN samples concentration of graphitic N is increased.

According to our XPS analysis, nitrogen in the RT sample primarily exists in form of N-C species that are common for nitrogen present in a carbon framework, with additional possible contributions from $FeN_x$ complexes. The majority of iron is found in oxide form, with the carbon also showing some oxidation. In the HT sample, iron nitride species are formed, in addition to the same N-C, $FeN_x$, and iron oxide species present in RT. Additionally, more graphitization of the carbon matrix was observed in HT sample due to the high temperature deposition (Figure S1c). The composition of the AN sample presents an intermediate case relative to the RT and HT samples, with fewer iron nitrides, an intermediate amount of N-C and $FeN_x$ species, and the highest relative amount of graphitic nitrogen, as well as hydrogenated nitrogen, amongst the three samples.

The increase in graphitization observed for the HT and AN samples has significant implications towards ORR performance, as the nitrogen defects that are stabilized within a graphitic carbon network are more likely to be active and stable than nitrogen species in a more disordered carbon. Additionally, this likely explains the increased electronic conductivity observed for HT and AN samples, deduced from Figure 3a. From the combination of the XPS analysis and RDE testing, it is possible to conclude that all three samples in as-prepared state contain species that have been attributed to ORR activity – pyridinic nitrogen, $Fe-N_x$ complexes, hydrogenated nitrogen (in-plane and edge conformation), and graphitic nitrogen, all in different proportions.

The results of XPS analysis post-RDE testing are shown for the N 1s in Figure 5 and the analysis results are summarized in Table 1. Due to the significant changes in relative N amount, we normalized data to the total N 1s area in Figure 5, so that relative comparisons can be made. As expected from exposure to $H_2SO_4$ electrolyte, all metallic iron, iron oxide, and iron nitride species are removed post RDE. Any possible Fe species remaining are below the detectable limit of the XPS instrument (<0.1 at. %), as shown in Table 1. In addition, significant changes were observed in N amount, with the HT and AN samples showing higher N concentration than the RT sample. All samples also show a significant shift in N 1s signal from lower BE to higher BE post-RDE testing. The position of the N 1s core level for HT and AN is in good agreement with protonated nitrogen species, and graphitic nitrogen, while the RT sample also shows the formation of a significant shoulder at this position not present in the as-produced film.[28] This increase in high BE signal in post-RDE testing N 1s spectra for RT films is accompanied by a corresponding loss in lower BE signal, suggesting the protonation of an electron-rich species, likely pyridine, is responsible for the shift in BE.

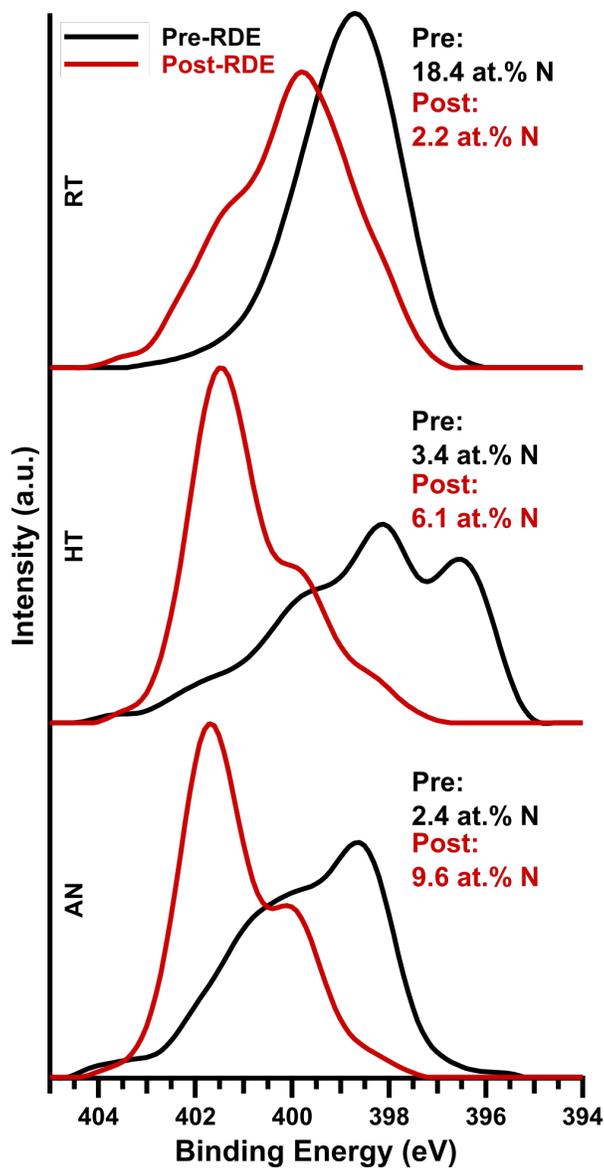

Figure 5: Overlays of representative background subtracted, area normalized N 1s spectra collected before and after RDE testing. For all samples, a significant shift to higher binding energy is observed after RDE testing. Since spectra are normalized to the total peak area to account for differences in N amount, at. % of N values from Table 1 are presented.

A complete analysis of compositional changes and XPS core level plots for the O 1s and C 1s can be found in Figure S2. All 3 samples have their primary C 1s peak between 284.4 and 284.7 eV, indicating a high amount of graphitic carbon, and suggesting that graphitic species are the most stable carbon species under ORR conditions, as expected. The HT and AN samples show a strong secondary peak at 286.1 -286.3 eV, that can be attributed to C-O, C-N species, or both. An increase in oxidation of the carbon matrix, indicated by O 1s spectra, could be due to the generation of a peroxide intermediate during the stepwise 2x2 e⁻ pathway of the ORR.[21, 29, 30] This feature therefore can be assigned to either oxidized carbon species due to the production of peroxide, or stabilized nitrogen defects within the carbon matrix. ORR activity trends with the relative intensity of this component, increasing from RT to HT samples and having a maximum value for AN sample. Additionally, the trend in relative abundance of this species post-RDE testing with relative nitrogen amount remaining in the film suggests that the stabilization of nitrogen-based defects within a graphitic carbon network during high temperature processing may result in available ORR active sites, most likely either graphitic nitrogen, hydrogenated nitrogen, or $FeN_x$ complexes.

### *3.3 Correlation of ORR and XPS results*

The XPS and ORR results presented above indicate that the ORR active $Fe-N_x$ and N-C moieties most likely are stabilized during high temperature growth, or upon annealing of Fe-N-C precursors deposited at RT. In these Fe-N-C precursors prepared by room temperature deposition, on the other hand, the elements most likely exist in mixed iron oxide and disordered N-C forms. XPS characterization also shows that high temperature exposure increased the relative abundances of graphitic-N and hydrogenated-N moieties in the annealed HT and AN samples. However, the HT sample still has a higher relative concentration of inactive Fe-oxides or Fe-nitrides, whereas the AN sample has the highest relative abundance of graphitic and hydrogenated-N moieties. Thus, while higher current densities and onset potentials obtained for HT and AN (Figure 3), correlate with the highest relative abundance of graphitic and hydrogenated nitrogen species (Figure 4), changes in chemical composition (Figure 5) and surface morphology (Figure 2) may be still influencing this correlation.

Table 2: XPS parameters found to trend with the current densities at 0.2 V, 0.3 V and 0.4 V of RT, HT and AN samples from RDE measurements. Relative contribution to the N 1s fit of the as-prepared films for three components (graphitic, and both in-plane and edge conformation hydrogenated-N) are summed, and displayed along with post-RDE N content (at.%).

| Thin Film Samples | XPS: Pre-RDE Testing | XPS: Post-RDE Testing | ORR Current Density, J (mA/cm²) | | |
|---|---|---|---|---|---|
| | % of all graphitized species in N 1s peak | N Relative Amount (at.%) | 0.4 V | 0.3 V | 0.2 V |
| RT | 9.4 | 2.2 | -0.009 | -0.005 | -0.007 |
| HT | 16.3 | 6.1 | -0.030 | -0.103 | -0.214 |
| AN | 35.8 | 9.6 | -0.142 | -0.362 | -0.600 |

The correlation between the ORR activity (represented by current density at certain voltage) and the concentration of different N-containing surface chemical species both pre RDE and post RDE testing is summarized in Table 2. Since a multitude of specific nitrogen and carbon moieties (Figure 4 and Figure 5) with different concentrations (Table 1 and Table S1) was shown to form depending on the processing temperature and other conditions, Table 2 lists only several parameters which trend with the ORR activity (Figure 3). The ORR activity in Table 2 is represented by the current densities obtained from RDE measurements at several different voltages, due to thin film geometry of the experiment. As shown in Table 2, the ORR activity increases as the fraction of the graphitized species in N 1s peak in pre-RDE measurement increases, and as the relative amount of N in post-RDE measurement increases. We stress that based on the data available in this paper, the relationship of these species with the ORR activity presented in Table 2 is correlative and not necessarily causal. Further experiments would be necessary to provide definitive, qualitative and quantitative information about the effect of various species and their concentrations on the ORR taking place in acidic media.

## 4. Conclusion

This study presents synthesis-structure-property correlations of chemical moieties present in sputtered Fe-N-C thin films as model systems for the ORR catalysts in acidic electrolytes. A set of model Fe-N-C ORR catalyst films was produced by co-sputtering iron and carbon onto glassy carbon substrates in a reactive nitrogen atmosphere at three different conditions – room temperature (RT) deposition, 650 °C deposition (HT), and RT deposition followed by a 750 °C anneal in nitrogen (AN). Deposition of each film onto a removable glassy carbon RDE tip enabled electrochemical testing through CV and LSV techniques to evaluate ORR activity. Characterization was performed by SEM to evaluate the sample morphology, while the relative abundances of various nitrogen species in the Fe-N-C thin films were evaluated with XPS, which was performed both pre RDE and post RDE tests. These results were correlated with the ORR activity.

SEM results indicate that depositions at elevated temperature (HT sample) can produce smoother films than the room-temperature deposition (RT sample) followed by the annealing (AN sample). XPS analysis results showed that a set of Fe-N-C materials with varying concentrations and speciation of iron, nitrogen, and carbon were produced at different processing conditions (i.e. RT, HT, AN). In particular, the types and the concentrations of the nitrogen moieties - graphitic-N, $FeN_x$ complexes, hydrogenated-N (representing both pyrrolic nitrogen and hydrogenated pyridine), and pyridinic-N - varied considerably, with some of these species correlating with ORR activity. Specifically, the improvement in ORR activity of the Fe-N-C thin films correlates with an increase in graphitic and hydrogenated-N moieties in the as-synthesized thin films. Moreover, high temperatures result in the formation of iron nitrides and possibly Fe-N$_x$ active sites, along with graphitization of the carbon within the film.

ORR testing followed by XPS measurements showed that the remaining nitrogen concentration in the films increases for samples exposed to elevated temperatures. The changes in the film composition post-RDE testing are primarily due to protonation of pyridinic nitrogen species under ORR conditions, while all iron containing species are leached out of the film due to their low stability in acidic environment. From the comparison of the ORR and XPS results, we conclude that high temperature processing that promotes graphitization of the Fe-N-C films is necessary to produce and/or stabilize active ORR sites accessible for catalysis. Optimization of the sputtering technique is necessary to quantitatively correlate the active sites and activity, and to progress towards derivation of causal relationships between them. Overall, this study offers a method for synthesis of model samples with more complex chemistries and smoother morphology that are typically reported. This method may assist in elucidating the role of active site moieties in Fe-N-C thin films, contributing to development of non-PGM catalysts for ORR in PEM fuel cells, and for OER in PEM electrolyzers.

**Acknowledgement**

This work was authored by the National Renewable Energy Laboratory (NREL), operated by Alliance for Sustainable Energy LLC, for the U.S. Department of Energy (DOE) under contract no. DE-AC36-08GO28308. Funding provided by the Office of Energy Efficiency and Renewable Energy (EERE), under Fuel Cell Technologies Office (FCTO), as a part of ElectroCat and HydroGEN Energy Materials Network (EMN) consortia. XPS work by M.J.D and S.P. was supported by start-up funds from Colorado School of Mines, and NSF Award #1800585 - Probing Catalyst-support Interactions via Experiment and Theory. Dr. Glenn Teeter is gratefully acknowledged for providing access and maintenance to XPS characterization facilities at NREL. The views expressed in the article do not necessarily represent the views of the DOE or the U.S. Government.

Supporting information

**Oxygen Reduction Reaction and X-ray Photoelectron Spectroscopy of Sputtered Fe-N-C Films**

Yun Xu[a], Michael J. Dzara[b], Sadia Kabir[a]. Svitlana Pylypenko[a,b*], Kenneth Neyerlin[a*], Andriy Zakutayev[a,*]

[a] Materials and Chemical Science and Technology Directorate, National Renewable Energy Laboratory, Golden, CO 80401, USA.

[b] Chemistry Department, Colorado School of Mines, Golden, CO 80401

*Corresponding authors: Andriy.Zakutayev@nrel.gov, Kenneth.Neyerlin@nrel.gov, spylypen@mines.edu

**1. Development of N 1s Peak Fit**

All components were constrained to a 0.1 eV range in position, with a set value for FWHM to ensure a consistent fit. The FWHM must be constrained to ensure a narrow peak – 1.0 eV is used in the case when only one type of species are likely to contribute at that binding energy, while broader peaks (1.3 - 1.5 eV) are used when a component cannot be assigned to a single species. Fitting parameters and relative abundance of each component can be found in SI Table1. Assignment of components to physical species is based upon a combination of prior experience working with organic nitrogen species and other metal nitrides, along with survey of the literature (citations in text at first mention of Figure 3). The two lowest BE components are assigned to iron nitrides of differing stoichiometry, with the lowest BE species likely to be a more nitridic species, and the higher BE component likely to be a more Fe rich stoichiometry. A variety of species rich in nitrogen such as inorganic carbon-nitride, N clusters, triazines, and imines can contribute to the peak centered at 398.0 eV. Pyridine is assigned to the component at 398.7 - 398.8 eV. Several species are possible for the peak at 399.5 - 399.6 eV, with amines and Fe-coordinated to organic N complexes contributing to this component. Two overlapping species make up the peak at 400.6 - 400.7 eV; pyrrolic nitrogen and hydrogenated pyridine present in an in-plane conformation. Graphitic nitrogen species, which can be classified as ordered/disordered depending on the presence of other defects which perturb the electronic structure of the graphitic nitrogen atom, are assigned to the broad peak at 401.8 - 401.9 eV. The final component at 403.5 – 403.6 eV is assigned to edge-conformation hydrogenated nitrogen, although it should be noted that oxidized organic-nitrogen species such as pyridine oxide and amide species can contribute at this and higher binding energies.

## N 1s Fitting Parameters

| Sample | Fe-Nitride 1 | | | Fe-Nitride 2 | | | N-rich | | | Pyridine | | |
|---|---|---|---|---|---|---|---|---|---|---|---|---|
| | BE | FWHM | % | BE | FWHM | % | BE | FWHM | % | BE | FWHM | % |
| RT | X | X | 0.0 | X | X | 0.0 | 398.0 | 1.4 | 36.9 | 398.8 | 1.0 | 18.5 |
| HT | 396.0 | 1.0 | 10.9 | 396.7 | 1.0 | 15.5 | 398.0 | 1.4 | 33.4 | 398.8 | 1.0 | 4.0 |
| AN | 396.0 | 1.0 | 0.6 | 396.7 | 1.0 | 0.4 | 398.0 | 1.4 | 18.1 | 398.7 | 1.0 | 17.7 |

| Sample | Amine/FeNx | | | Hydrogenated N (in-plane) | | | Graphitic | | | Hydrogenated N (edge) | | |
|---|---|---|---|---|---|---|---|---|---|---|---|---|
| | BE | FWHM | % | BE | FWHM | % | BE | FWHM | % | BE | FWHM | % |
| RT | 399.5 | 1.4 | 35.2 | 400.6 | 1.4 | 6.8 | 401.8 | 1.5 | 2.6 | X | X | 0.0 |
| HT | 399.6 | 1.4 | 19.9 | 400.6 | 1.4 | 8.1 | 401.9 | 1.5 | 7.1 | 403.5 | 1.3 | 1.1 |
| AN | 399.6 | 1.4 | 27.5 | 400.7 | 1.4 | 22.2 | 401.8 | 1.5 | 11.2 | 403.6 | 1.3 | 2.4 |

*Table S1:* *Fitting parameters (BE, FWHM, and relative percentage of the fit) of the N 1s are displayed for all components of all as-prepared samples. Components that are not present in a sample are denoted with an X in the BE column.*

## 2. XPS of As-prepared Fe-N-C Films

Compositional analysis of XP spectra for the Fe 2p, O 1s, and C 1s are shown in Figure S1. The RT sample shows features of mixed iron oxides, and oxidized carbon species, while the HT sample has additional feature of iron nitrides and likely metallic Fe. The Fe $2p_{3/2}$ region in Figure S1a has an additional feature not present in the RT sample, a small low BE shoulder. This reduced state present at 706.9 eV is most likely metallic Fe, which has been assigned to a range of values between 706.5 and 707 eV.[1-2] It is possible that iron nitride contributes to this feature, however it is more likely to appear at a BE between that of metallic iron and iron oxide, as coordination with nitrogen is likely to result in an iron species with less electron density than that in a metallic structure, but more than that of iron in an oxide. Indeed, at slightly higher BE values than that attributed to metallic iron, there is a slight asymmetric broadening to lower BE that is present in the Fe $2p_{3/2}$ at 707-708 eV not observed in the RT sample, which can be attributed to the formation of an iron rich nitride phase.[3] The increase in relative O and decrease in relative N content reported in Table 1 can be explained by the changes in O 1s and C 1s peak shapes relative to the RT sample as shown in Figure S1b and Figure S1c. The O 1s peak max is at 530.0 eV, showing that oxygen present in iron oxide(s) is now the dominant oxygen species. The C 1s shows less signal from 285 - 288 eV, which is the result of less oxidation and nitridation of carbon. What is also worth noting is the shift in the C 1s peak max from 284.8 eV to 284.5 eV, which indicates an increased graphitization of the carbon matrix.[4-5] This corroborates the results from Table 1, as an increase in carbon content relative to the total amount of oxygen and nitrogen is expected when comparing a graphitic system (HT) to a more disordered, defected system (RT).

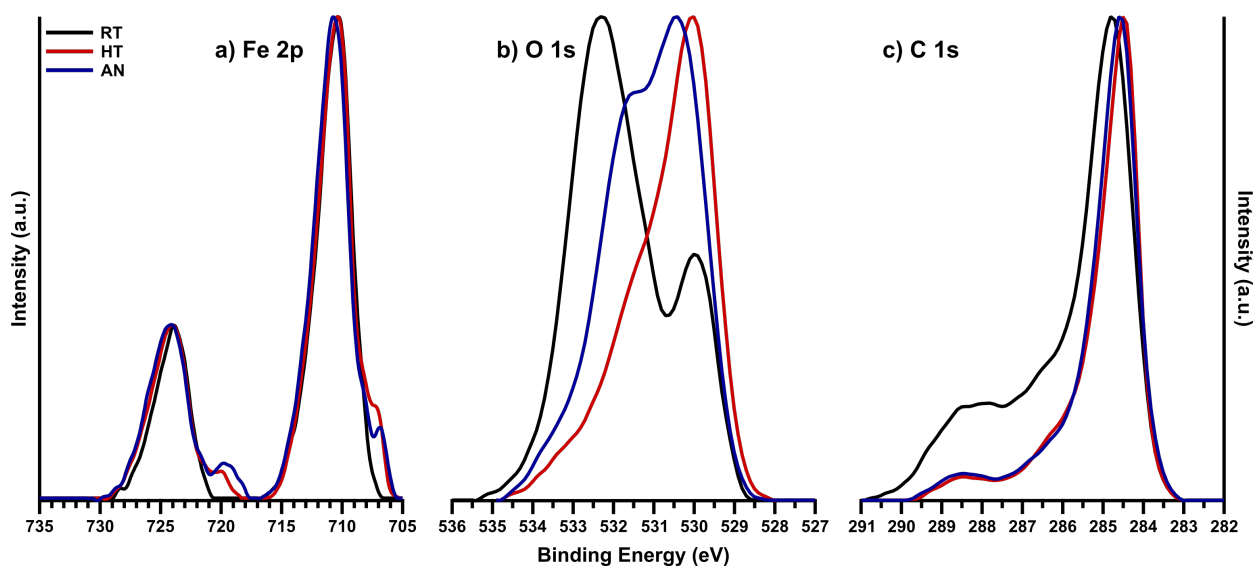

*Figure S1. Background subtracted, peak max normalized spectra are shown overlaid for each sample for the relevant core levels – a) Fe 2p, b) O 1s, c) C 1s.*

With respect to nitride formation, the composition of AN sample presents an intermediate case relative to the RT and HT samples. Nitrogen is present primarily as organic functionalities, with the highest relative amount of graphitic nitrogen, as well as hydrogenated nitrogen, among the three samples. While some nitrides are formed, it is at a much lower concentration than the HT sample. Despite this, the Fe $2p_{3/2}$ feature at 706.9 eV has a similar relative intensity as the HT sample, further indicating that this feature is likely due to the presence of a small amount of metallic iron. This feature is better resolved from the slight broadening at 707 - 708 eV than in the HT, consistent with the N 1s results that suggest some nitride is present in the AN sample, but at a lower amount than in the HT. The O 1s and the C 1s for the AN sample are very similar to that of the HT sample, with the only significant difference present in the relative contribution of iron oxide(s) and oxidized carbon to the O 1s – the AN sample has a more significant contribution from oxidized carbon.

**3. XPS Post-RDE Testing**

O 1s and C 1s core levels are plotted overlaid for each sample in Figure S2. The loss of nitride species is observable from the N 1s, while the loss of oxides is confirmed in the O 1s (Figure S2a) for all samples, as there is little to no signal from 529-530 eV. Within the N 1s, the RT sample has slightly different characteristics than the two films processed at higher temperatures. All samples show a significant shift in N 1s signal from lower BE to higher BE. It is possible that some of the signal loss is due to delamination or dissolution during testing, particularly in the case of the RT sample where the relative nitrogen amount decreased significantly post testing. The relative increase in nitrogen amount for the HT and AN samples, likely due in part to the loss of iron oxide species, indicates that a process other than delamination/dissolution is responsible for the changes in the N 1s features of these samples post RDE testing. Further insights into material changes and possible stable active sites can be obtained by examining the changes in peak shape for each sample.

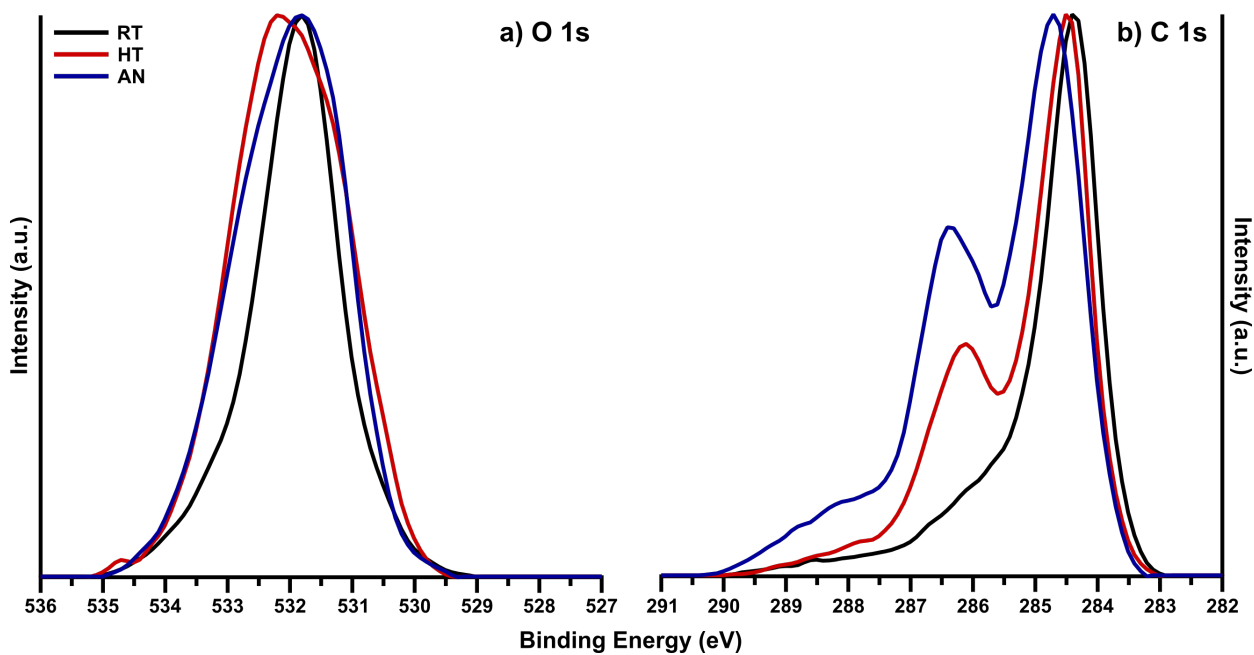

*Figure S2. Background subtracted, peak max normalized spectra collected on RDE tips post-ORR testing are shown overlaid for each sample for the relevant core levels – a) O 1s, and b) C 1s*

The position of the N 1s peak maximum is at 399.8 eV for the RT sample, while HT and AN have values that are shifted to higher BE - 401.5 and 401.7 eV, respectively. For the RT sample, this value corresponds closest to $FeN_x$ complexes, and hydrogenated nitrogen species. The position for HT and AN is in good agreement with protonated nitrogen species, and graphitic nitrogen, while the RT sample also shows the formation of a significant shoulder at this position not present in the as-produced film.[6] This increase in high BE signal in post-ORR in N 1s spectra is accompanied by a corresponding loss in lower BE signal, suggesting the protonation of an electron-rich species is responsible for the shift in BE. Indeed, it has been previously shown that pyridinic nitrogen is readily protonated under the conditions necessary for the ORR in acidic media.[6-7] A small amount of signal remains at 398-399 eV, indicating that some pyridinic nitrogen remains in the film post-ORR testing, however it is likely in the depth of the film where it would not be accessible to the corrosive electrolyte and therefore is not protonated. Signal remains at the position of each other possible active species post-ORR testing, however the overlap between protonated and graphitic nitrogen prevents accurate comparisons of the relative abundance of each species.

Analysis of the C 1s (Figure S2b) further confirms the nature of the differences seen in N 1s changes among the 3 samples post-ORR testing. All 3 samples have their primary peak between 284.4 and 284.7 eV, indicating a high amount of graphitic carbon, and suggesting that graphitic species are the most stable carbon species under ORR conditions. This also explains the dramatic change in nitrogen amount for the RT sample – the majority of its nitrogen species were likely formed in disordered carbon that was not stable during ORR testing. It is possible that the shift to lower BE observed for RT is due in part to a higher contribution from the glassy carbon substrate if a significant amount of the film dissolved or delaminated. Within the post-ORR C 1s, there is also a feature of note from 286.1 - 286.3 eV, which appears as a shoulder in the case of the RT sample, and a more well resolved peak for the two high temperature samples. Several species are possible at this BE, including C-O-C and C-OH species, as well as nitrogen defects in graphitic carbon.[8] The possible contributions of C-O-C and C-OH species, and the very small shifts in C 1s position of the different nitrogen defects prevent this feature from being unequivocally assigned to any one Nx or C-O species.

**Supporting References**